\documentclass{aa} 
\input psfig.sty

\begin{document}


%
   \title{Multisite observations of the PMS $\delta$ Scuti star 
V351 Ori\thanks{Based on 
observations collected at the: Loiano Observatory (Italy),  
San Pedro Martir Observatory (M\'exico), OGS at Teide Observatory (Spain), 
JKT at Roque de los muchachos Observatory (Spain), 
Beijing Astronomical Observatory (China), 
SARA Observatory at Kitt Peak, which is owned and operated 
by the Southeastern Association for Research in Astronomy.}}


\author{V. Ripepi\inst{1}, M. Marconi\inst{1}, S. Bernabei\inst{2,3}, 
 F. Palla\inst{4}, F. J. G. Pinheiro\inst{5},  D. F. M. Folha\inst{5}, 
T.D. Oswalt\inst{6}, L. Terranegra\inst{1}, A. Arellano Ferro\inst{7},  
X.J. Jiang\inst{8}, J.M. Alcal\'a\inst{1}, S. Marinoni\inst{2},
M.J.P.F.G. Monteiro\inst{5}, M. Rudkin\inst{6}, K. Johnston\inst{6}}

\offprints{ripepi@na.astro.it}

\institute{INAF-Osservatorio Astronomico di
Capodimonte, Via Moiariello 16, I-80131 Napoli, Italy \and
INAF-Osservatorio Astronomico di Bologna, Via Ranzani 1, 
40127 Bologna, Italy \and
Departimento de Astrof\'{\i}sica, Universidad de La Laguna, Avda. 
Astrofisico F. S\'anchez sn, 30071 La Laguna, Spain \and 
INAF-Osservatorio Astrofisico di Arcetri, Largo E. Fermi, 5, I-50125
Firenze, Italy \and
Centro de Astrof\'{\i}sica da Universidade do Porto, Rua das Estrelas,
4150-762 Porto, Portugal \and
Florida Inst. Technology, 150 W Univ. Blvd., Melbourne, FL 32901-6988, USA \and
Instituto de Astronom{\'\i}a, UNAM, Apdo. Postal 70-264, M\'exico D.F.,
CP 04510, M\'exico \and 
National Astronomical Observatories, Chinese Academy of Sciences, 
Beijing, 100012, China \\
}

 \date{Received/Accepted}

\abstract{We present the results of  multisite observations 
spanning two years on the pre--main-sequence (PMS) star V351 Ori.  
A total of around 180 hours of observations over 29 nights have been 
collected, allowing us to measure five different periodicities, 
 most likely related to the $\delta$ Scuti variability of V351 Ori.
Comparison with the predictions of linear
nonadiabatic radial pulsation models put stringent constraints on the
stellar parameters and
indicate  that the distance
to V351 Ori is intermediate between the lower limit measured by Hipparcos
(210 pc)
 and that of the Orion Nebula (450 pc). However, radial pulsation models are 
unable to reproduce all of the observed frequencies with a single choice
of (M$_\ast$, L$_\ast$, and T$_{\rm eff}$), suggesting 
the presence of additional nonradial modes.

}

\titlerunning{Multisite observations of the PMS $\delta$ Scuti star V351 Ori}
\authorrunning{Ripepi et al.}

   \maketitle

%

\section{Introduction}
\label{intro}

$\delta$ Scuti stars are usually associated with the main-sequence (MS) or 
post-MS evolutionary phases.  To date, about ten intermediate-mass
pre--main-sequence (PMS) stars have been found to pulsate with timescales
typical of $\delta$ Scuti variables and a growing interest in these young
objects has developed in the last few years (see Marconi et al. 2001,
hereafter M01; Marconi, Palla \& Ripepi 2002, hereafter M02, and references
therein).  The first two candidates were discovered by Breger (1972) in the
young cluster NGC 2264. The existence of PMS $\delta$ Scuti stars was later
confirmed by observations of the short term variability of the Herbig Ae
stars HR~5999 (Kurtz \& Marang 1995)  and HD~104237 (Donati et al. 1997).

\par
The initial evidence has stimulated the theoretical investigation of the PMS
instability strip, based on nonlinear convective hydrodynamical models
(Marconi \& Palla 1998). As a result, the topology of the instability strip
for the first three radial modes was identified and a list of possible PMS
pulsating candidates, mainly Herbig Ae stars, was selected on the basis of
their spectral types.  Then, a series of observing runs has been carried out
searching for $\delta$ Scuti-type photometric variations with periods of
minutes to a few hours and amplitudes less than a few tenths of a
magnitude among those stars inside or close to the theoretical instability
strip.
These studies have confirmed the $\delta$ Scuti variability
of HD~104237 (Kurtz \& Muller 1999), HR~5999 (Kurtz \& Catala 2001), and six
additional candidates have been identified (see Kurtz \& Muller 2001; M01;
M02).

Among the new discoveries, V351~Ori is a particularly interesting object.
The star is characterized by strong long-term photometric variations,
attributed to extinction by circumstellar dust clouds. This property has been
considered a sign of youth, signaling the transition from an active PMS phase
to the arrival on the main sequence. Similar behavior has been observed in
other well known Herbig Ae stars, such as BN Ori, LkH$\alpha$~234 and UX Ori
(e.g., Herbst \& Shevchenko 1998).  The spectral properties of V351~Ori
include the presence of H$\alpha$ emission and a pronounced inverse P Cygni
profile (van den Ancker et al. 1996), the SI~$\lambda$1296~\AA\ and
OI~$\lambda$1304~\AA\ ultraviolet lines (Valenti et al. 2000), and a broad
infrared excess (van den Ancker et al. 1997). Altogether, these features
clearly reveal an active phase of interaction with the circumstellar
environment.  On the other hand, the possibility that V351~Ori is a post-MS
star has been put forward in the literature.  Evidence in this sense comes
from the lack of a reflection nebulosity, the absence of dense molecular gas
around the star, and a space velocity which indicates that the star is
approaching Orion and hence is not kinematically linked to the star forming
region. Spectral analysis of hydrogen absorption lines also show some
peculiarities (Koval'chuck \& Pugach  1998; see also M01 for details).
Recently, on the basis of high dispersion echelle spectra, 
Balona, Koen \& van Wyk (2002, hereafter BKW) 
have provided further evidence that V351~Ori is indeed a PMS star.

As already remarked, the interest for V351~Ori lies in the fact
that it pulsates. Marconi et al. (2000) and M01 have interpreted the
frequencies observed in the light curve of V351 Ori as due to
pulsations in a mixture of several radial modes. In particular, in M01 four
of the six observed frequencies have been simultaneously reproduced by a
linear nonadiabatic pulsation model with mass $M_{\ast} = 1.8\, M_{\odot}$,
luminosity $L_{\ast} = 13.87\, L_{\odot}$ and effective temperature $T_{\rm
eff} = 7350\, K$.  However, the comparison of the observed periods with 
model predictions could not help
distinguishing whether the star is in the PMS or post-MS evolutionary phase of 
its evolution, even if the
solution is consistent with the distance determined by Hipparcos. The latter,
if correct, implies that V351~Ori is much closer than the Orion complex.

Owing to the short temporal coverage of our previous observations, the
published periodicities are subject to the problem of one cycle per day
aliasing, rendering their {\b determination} rather uncertain. With the aim
of overcoming this observational limitation, we have organized a multisite
campaign on V351~Ori involving telescopes located in five different
countries.  In this paper, we present the results of this campaign. The
organization of the paper is the following: in Section 2 we present details
on observations and data reduction; the frequency analysis is reported in
Section 3 and results are compared with those obtained in other works and
with the predictions of pulsating models in Section 4; the conclusions
close the paper.

\section{Observations and data reduction}
\label{obsred}

Six observatories were involved in the observational 
campaign.
They are listed in Table~\ref{obs} together 
with the telescopes and instruments used. The observations have been 
supplemented by data obtained at the SAAO during the same period (see BKW). 
Two comparison stars (HD~38248 and HD~290826) have been used during the 
observations, depending on the constraints set by the different instruments 
(e.g by the field of view of the CCD etc.). 
Some information on these objects is given in 
Table~\ref{comparison}, along with the same data for V351~Ori. 
For completeness, the last column of Table~\ref{obs} indicates the comparison 
star used at each telescope. 

\begin{table*}
\caption[]{List of telescopes and instruments involved in the
multisite campaign.} 
\label{obs}
\begin{tabular}{llll}
\hline
\noalign{\smallskip}
Observatory  & Telescope & Instrument & Comp. Star\\
\noalign{\smallskip}
\hline
\noalign{\smallskip}
SAAO (S.Africa)                & 0.5m     & Modular Photometer & HD~290826 \\
Roque de los Muchachos (Spain) & 1.0m JKT & SITe2 2048x2048 CCD & HD~290826 \\
Beijing Astronomical Observatory - BAO (China) & 0.85m & Three Channel Photometer & HD~290826 \\
Loiano (Italy)    & 1.5m  &  Three Channel Photometer  (TTCP) & HD~290826 \\
San Pedro Martir - SPM (Mexico) & 1.5m  & Danish ($uvby$) Photometer & HD~290826 \\
Kitt Peak National Observatory (USA)  & 0.9m SARA & 512x512 CCD & HD~38248  \\
Teide (Spain)     & 1.0m OGS & 1024x1024 CCD & HD~38248 \\
\noalign{\smallskip}
\hline
\end{tabular}
\end{table*}

The multisite observations on V351~Ori were performed in the 
Johnson $V$ filter  over around 2 years. 
We can divide the campaigns in: 1) winter 2001 
(Sept. 2001-Feb. 2002); 2) winter 2002 (Oct. - Dec. 2002).
In total, we gathered about 180 hour of observations over 29 nights.  
The detailed log of the observations is given in Table~\ref{jou}.


\subsection{Details on single site observations}

The observations obtained at SAAO are described in
BKW. Considering the BAO and Loiano observations, in both cases the
instrument was a three channel photometer (see www.bao.ac.cn for the BAO
instrument and www.na.astro.it/$\sim$silvotti/TTCP.html for the TTCP in
Loiano).  These instruments allow the simultaneous observation of 
the variable and comparison stars, as well as the sky. 
Thus, the time sampling can be very
short: for V351~Ori we adopted an exposure time of 10 s.


\begin{figure}
\caption{Comparison between data collected during the same night 
(November 29, HJD=2452243) at BAO (crosses), SAAO (open circles), and 
at JKT (filled circles).} 
\label{confr_vari}
\psfig{figure=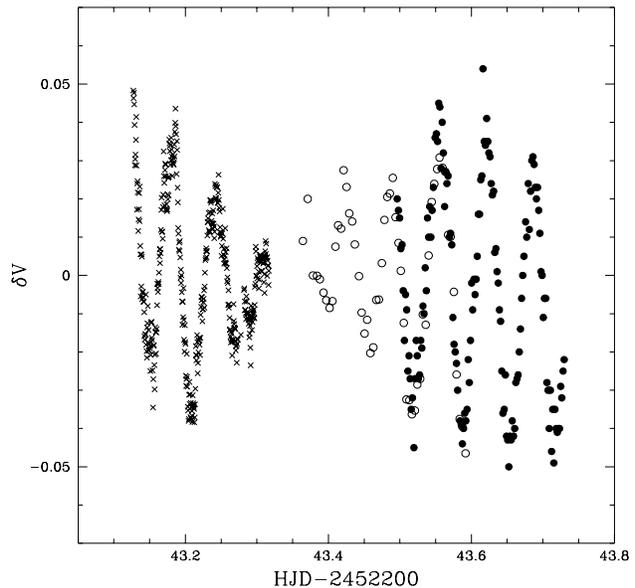,width=8.5cm}
\end{figure}

Observations in SPM were carried out using the six channel spectrophotometer 
(see 132.248.3.38/ Instruments/danes/photomdan.html)
which can simultaneously operate the $uvby$ filters. For V351 Ori, we 
used only the $y$-band data after the proper conversion into the Johnson $V$
band. 

Concerning CCD observations, below we report the data acquisition and 
reduction in detail.

\subsection{Roque de los Muchachos Observations}

The observations with the 1.0m JKT (La Palma, Canary
Islands, Spain) were carried out during November-December 2001 
using the JAG CCD camera equipped with the
2048x2048 pix$^{2}$ SITe2 detector, providing a 11$\times$11 arcmin$^2$
field of view and a pixel scale of 0.33 arcsec/pixel. A Johnson $V$-band
filter was used during all observations with exposure times between 10 and
20 seconds (depending on airmass and seeing). When combined with the
read-out time the sampling rate is decreased to about one photometric point
every two and half minutes. Standard data reduction procedures were
performed with IRAF. Circular aperture photometry was carried out with
DAOPHOT under IRAF to find instrumental magnitudes of the simultaneously
observed target and comparison stars. The adopted aperture radius was 70
pixels (23.1 arcsec) for the night of November 29, and 25 pixels (8.25
arcsec) for the nights of November 29 and December 3.
Different aperture radii were used to account for the different 
seeing conditions.

\subsection{Kitt Peak Observations}
$V$-band observations were obtained with the SARA 0.9m automated
telescope at Kitt Peak in Arizona.  A little more than 42 hours
of differential time-series photometry was obtained on nine nights
between October 2001 and November 2002 by both on-site and remote-access
observers.  Images were collected using an Apogee AP7p camera with a
back-illuminated SITe SIA 502AB 512x512 pixel CCD.  The pixels 
are 24 microns square, corresponding to 0.73" at the telescope
focal plane scale.  Read out noise and gain for the camera are about 12.2 
electrons (rms) and 6.1 electrons/ADU, respectively.  

Integration times were typically 10-30 seconds (depending upon sky
conditions and airmass), plus $\sim$ 5 seconds readout and idle 
time between frames.
Sky flats, dark and bias exposures were taken every night.  All data
was calibrated and reduced using standard IRAF routines.

\subsection{Teide Observations}
Observations at Teide Observatory were carried out with the 
 1.0m OGS Telescope equipped with a 1024x1024 pixel CCD. 
The pixel scale was 0.32$^{''}$ for a total field of view 
of 5.5$^{'}$x5.5$^{'}$. Read noise and gain for the camera 
are about 5.4 electrons (rms) and 2 electrons/ADU, respectively.
Typical integration  times were 10-15 sec, depending upon sky conditions.
Sky flats, dark and bias exposures were taken every night.  All data
was calibrated and reduced using standard IRAF routines.


In order to homogenize the differential photometry obtained 
at telescopes which used different comparison stars and to overcome the 
consequent difficulty in obtaining a secure zero point stability, 
we have decided to subtract from the data gathered during each night 
at each site their mean values. This procedure will limit our 
analysis to frequencies longer than 1 c/d.  
However, since the longest time series obtained at a single 
site lasts around 7 h, all the frequencies 
lower than $\sim$ 3.3 c/d have to be considered suspect. 
However, this is not a serious drawback, since the range of frequencies 
of interest for $\delta$ Scuti pulsations generally falls beyond this value. 

In some case observations made at different observatories 
overlapped in time, in particular the run at SAAO and ING and/or Loiano. 
In this event, the data were averaged together. In order to compare 
the quality of the data from different sites taken at the same 
time, we show in Fig.~\ref{confr_vari} the observations collected 
at BAO, SAAO and ING during the night of November 29 (HJD=2452243).

In order to have an approximately similar sampling and a similar 
weight for each dataset in the Fourier analysis, we had to bin the 
photometry obtained at BAO, Loiano, OGS and SARA. 
As a consequence, the total number of data points has been reduced by a factor 
$\sim$4. In this way, we have the significant advantage of 
using much less computer time for frequency extraction.

We note that, in order to improve the analysis, and in 
particular the phase determination, 
we have added to the set of observations described above the data obtained 
about 1 year before and described in M01. We recall that
these observations were obtained at Loiano with the TTCP. These data
were binned in the same manner as described above.

\begin{figure*}
\psfig{figure=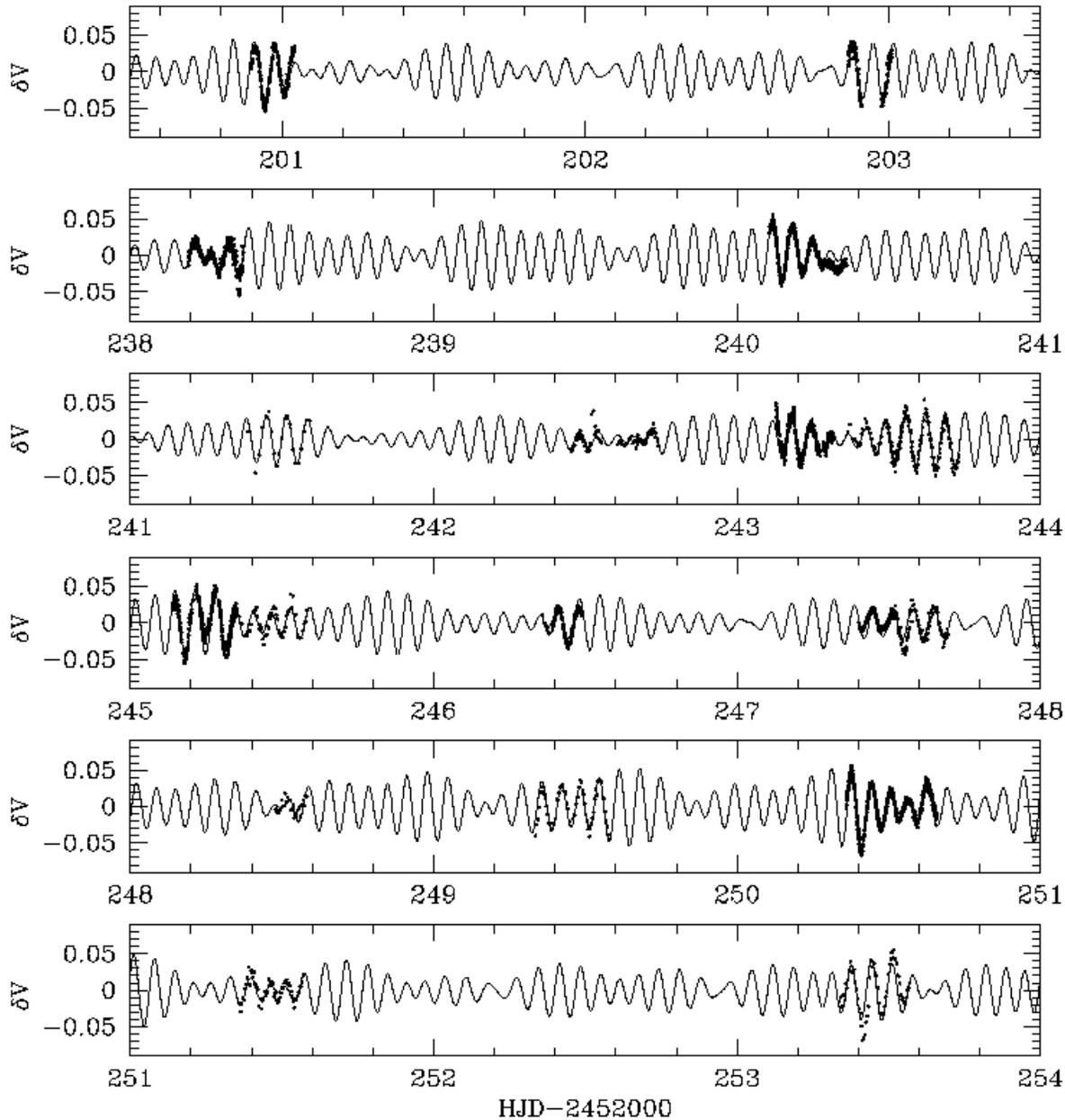,width=17cm}
\caption{Light curve of V351~Ori (points). The meaning of $\delta$V 
(in magnitudes) is Variable-Comparison star. The winter 2001 
and winter 2002 campaigns include data from 
HJD=2452201 to HJD=2452318 and from HJD=2452567 to HJD=2452586, 
respectively. The solid line shows the fit to the data 
obtained as described in Sect.~\ref{freq_res}}
\label{phot1}
\end{figure*}

\begin{figure*}
\psfig{figure=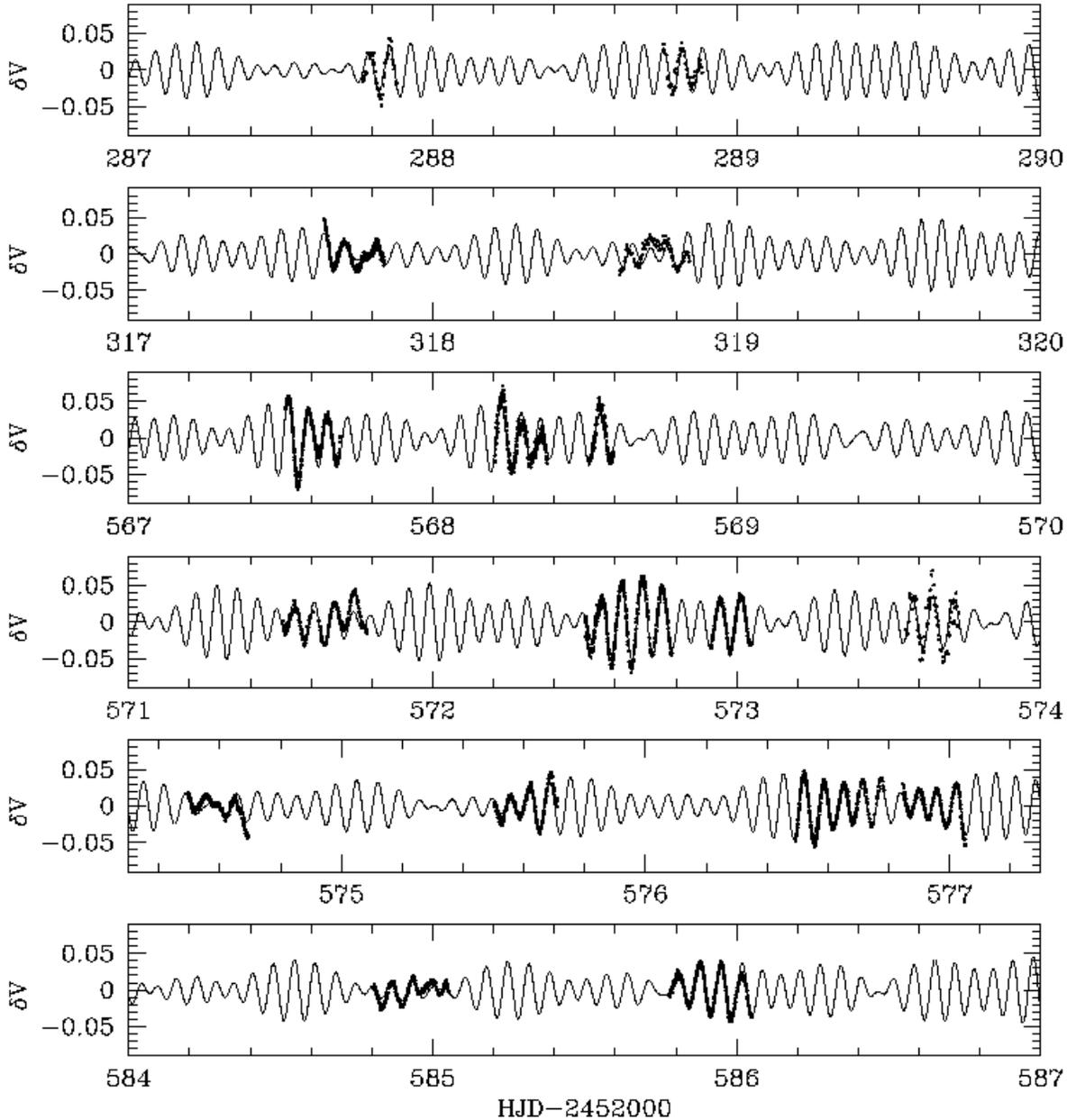,width=17cm}
\caption{Light curve of V351~Ori. Same as in Fig.~\ref{phot1}. } 
\label{phot2}
\end{figure*}

Finally, the light variations of V351 Ori obtained during the winter 2001 and 
winter 2002 campaigns are shown in Figs.~\ref{phot1},\ref{phot2}.


\begin{table}
\caption[]{Properties of V351~Ori and comparison stars.
\label{comparison}}
\begin{tabular}{ccccc}
\hline
\noalign{\smallskip}
Star  & AR      & DEC     & V     & Sp. T.   \\
      & (J2000) & (J2000) & (mag) &         \\
\noalign{\smallskip}
\hline
\noalign{\smallskip}
V351~Ori & 05 44 18.8 & +00 08 40.4 & 8.9  & A7  \\
HD~38248  & 05 44 23.8 & +00 05 08.2 & 9.0  & A3  \\
HD~290826 & 05 44 00.4 & +00 15 33.0 & 10.0 & F5  \\
\noalign{\smallskip}
\hline
\end{tabular}
\end{table}

\begin{table}
\caption[]{Journal of the observations.} 
\label{jou}
\begin{tabular}{ccl}
\hline
\noalign{\smallskip}
Starting HJD-2452000  & Duration & Observatory \\
   (day)    &   (hours)&  \\
\noalign{\smallskip}
\hline
\noalign{\smallskip}
200.88 & 3.5 & SARA \\
202.86 & 2.1 & SARA \\
238.19 & 4.4 & BAO \\
240.11 & 6.1 & BAO \\
241.39 & 4.9 & SAAO  \\
242.45 & 7.0 & SAAO+JKT \\
243.12 & 13.2 & BAO+SAAO+JKT \\
245.14 & 10.6 & BAO+SAAO \\
246.36 & 3.1 & SAAO+Loiano  \\
247.40 & 7.0 & Loiano+JKT \\
248.48 & 2.5 & SAAO \\
249.33 & 5.8 & SAAO \\
250.36 & 7.2 & SAAO+Loiano \\
251.36 & 5.2 & SAAO \\
253.34 & 5.6 & SAAO \\
287.76 & 2.9 & SPM \\
288.76 & 3.4 & SPM \\
317.64 & 4.8 & SARA \\
318.61 & 5.6 & SARA \\
567.51 & 4.4 & Loiano \\
568.20 & 6.1 & BAO+Loiano \\
571.50 & 6.4 & Loiano+OGS \\
572.50 & 10.2 & OGS+SARA \\
573.55 & 4.3 & OGS \\
574.50 & 4.7 & Loiano \\
575.50 & 5.0 & OGS  \\
576.49 & 12.0 & OGS+SARA \\
584.80 & 6.0 & SARA \\
585.78 & 6.5 & SARA \\
\noalign{\smallskip}
\hline
\end{tabular}
\end{table}

\section{Frequency analysis}
\label{freq_res}

\begin{figure}
\psfig{figure=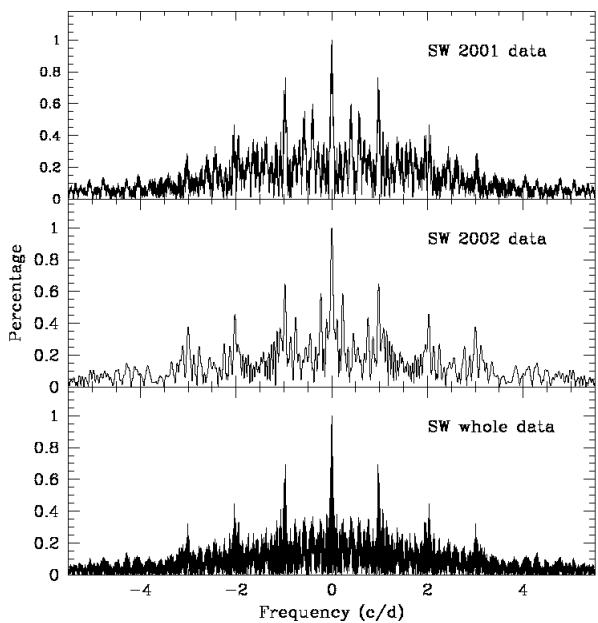,height=8.5cm}
\caption{Spectral window for the 2001 dataset (upper panel), 
2002 dataset (middle panel), and the whole dataset (lower panel). 
Note that the y-axis is in amplitude units.} 
\label{fig2}
\end{figure}

The frequency analysis was performed using the Period98 package
(Sperl 1998, www.astro.univie.ac.at/$\sim$dsn/), based on the Fourier
transform method. For a better interpretation of the results, we have
first calculated the spectral windows (SW) for the data set collected in 
winter 2001, winter 2002, and for the whole data set (which includes also 
the 2000 M01 data).  
The result is shown in Fig.~\ref{fig2}. 
As can be inferred from the figure, 
the 1 c/d alias is not significantly reduced when analysing 
the whole dataset with respect to the individual sets of data
(from $\sim$80\% to  $\sim$70\% in amplitude). However,  
 the main improvement obtained from analysing the whole dataset
is the strong decrease of aliases between 0 and 1 c/d.

\begin{table}
\caption[]{Frequencies, amplitudes and phases derived from the Fourier 
analysis of the data. Phases are calculated with respect to 
HJD=24552500.000 The uncertainty on the frequency is 
$\sim$0.03~c/d.
The last column lists the signal-to-noise ratio 
(in amplitude) calculated as described in the text 
(see Sect.~\ref{freq_res}). 
\label{tab2}}
\begin{tabular}{ccccc}
\hline
\noalign{\smallskip}
  & Frequency   &   Amplitude & Phase  & S/N \\
  & (c/d)       &    (mmag)   &        &     \\
\noalign{\smallskip}
\hline
\noalign{\smallskip}
$f_1$  &    15.687   &      22.9  &  0.759  & 6.8 \\
$f_2$  &    14.331   &      13.5  &  0.139  & 5.9 \\
$f_3$  &    12.754   &       8.4  &  0.750  & 5.7 \\ 
$f_4$  &    15.885   &       7.7  &  0.124  & 5.9 \\
$f_5$  &    12.817   &       4.8  &  0.196  & 4.6 \\
\noalign{\smallskip}
\hline
\end{tabular}
\end{table}

\begin{table}
\caption[]{Comparison between frequencies extracted from  the analysis 
of the winter 2001 and winter 2002 campaigns and the whole dataset.
The uncertainties on the frequencies for each data set are also indicated. 
\label{campagne}}
\begin{tabular}{ccccc}
\hline
\noalign{\smallskip}
  & whole dataset        & winter 2001          & winter 2002           \\
  & $\pm 0.03$ (c/d) & $\pm 0.06$ (c/d) &  $\pm 0.05$ (c/d) \\
\noalign{\smallskip}
\hline
\noalign{\smallskip}
$f_1$  &    15.687   & 15.687  & 15.685 \\
$f_2$  &    14.331   & 14.331  & 14.332 \\
$f_3$  &    12.754   & 12.756  & 12.771 \\ 
$f_4$  &    15.885   & 15.873  & 15.867 \\
$f_5$  &    12.817   &  --     & 12.797 \\
\noalign{\smallskip}
\hline
\end{tabular}
\end{table}

The spectral window is useful for estimating the error 
on the extracted frequencies. The latter is approximately  
equal to the FWHM of the main lobe in the spectral window 
(see, e.g., Alvarez et al. 1998). 
In our case, we find $\Delta f$ $\sim$0.03 c/d for the whole dataset, 
whereas the errors for the winter 2001 and winter 2002 campaigns are 
$\pm 0.06$~c/d and  $\pm 0.05$~c/d, respectively.

Before extracting the pulsation frequencies from the time series, 
it is important to have a tool to identify the last significant periodicity 
in the data set.
To this aim, we have adopted both the Scargle 
test (Scargle, 1982, see also Horne \& Baliunas 1986) and the  
empirical criterion suggested by Breger et al. (1993)
(see also Kuschnig et al. (1997)). These authors state that 
the signal-to-noise ratio (in amplitude) should be at least 
4 in order to ensure that the extracted frequency is  significant. 
In the following, we adopt this convention that will be referred to as 
``the Breger criterion''.

In order to apply the Scargle test, we have to estimate
the noise spectrum over the whole set of investigated frequencies and
the number of independent frequencies in the spectrum itself. 
To address the first point we have used the Period98 program 
to average the noise in boxes of 10 c/d width moving along 
the spectrum. The size of these boxes was chosen on the basis of the 
spectral window shape (see Fig.~\ref{fig2}) so that
at least part of the white noise was included. It is worth noting that,
as recommended by Horne \& Baliunas (1986), the noise spectrum used to 
investigate the significance  of a certain frequency $f_x$ is 
calculated {\it before} subtracting that frequency.
\par
The second point is more difficult to address, 
since there is no precise way of determining the number 
of independent frequencies in case of unevenly 
spaced data (which is our case). 
However, as discussed in Alvarez et al. (1998), we can use the FWHM of the 
main lobe in the spectral window ($\Delta f$ $\sim$ 0.03 c/d
for our dataset) to estimate this value. 
Using Eq. (20) of Scargle (1982), we find that 99\% and 90\% 
significance levels are reached when the signal 
(in amplitude) is 3.9 and 3.5 times the noise. 
Note that the Scargle Test is, in practice, equivalent 
to the Breger criterion\footnote{It is important 
to note that in Breger et al. (1993) the noise level is calculated 
{\it after} that {\it all} the relevant frequencies have been subtracted. 
To be conservative, we have adopted a noise level equal to the one used 
in Scargle's test.}. 
In order to rely only on secure frequency detection, 
we choose to consider only frequencies whose level of confidence  
is better than 99\% (i.e. S/N=3.9).

\begin{figure*}
\psfig{figure=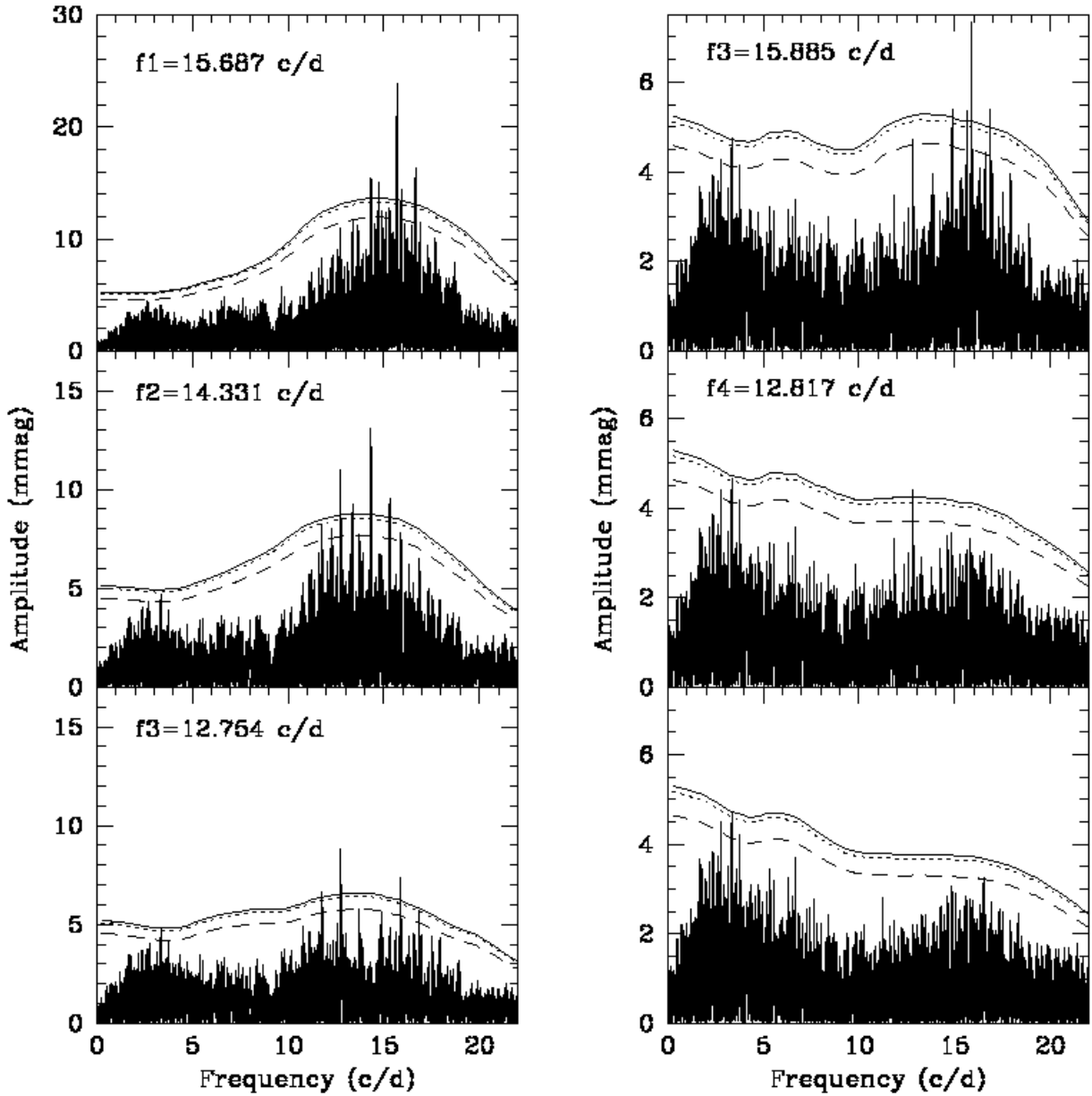,width=17cm}
\caption{Frequency analysis for the whole data set. Each panel 
shows the Fourier Transform after the subtraction of a pulsating 
The solid line corresponds to S/N=4.
The dotted and  dashed lines show the 99\% and 90\% significance 
level calculated from the Scargle (1982) test.} 
\label{fig3}
\end{figure*}
\begin{figure}
\psfig{figure=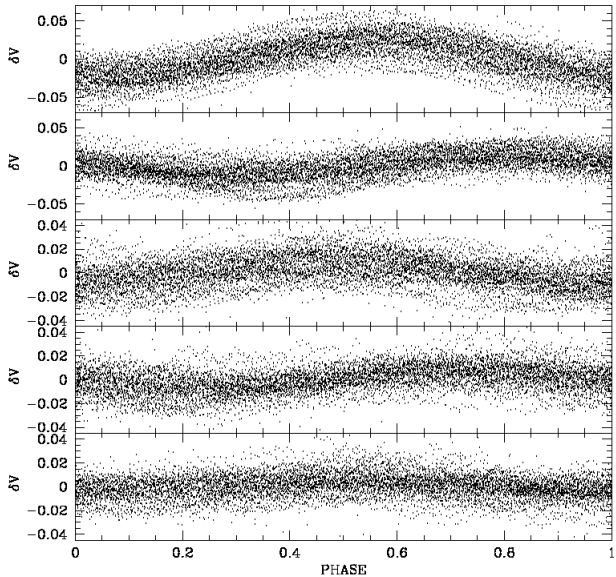,height=8.5cm}
\caption{Phase diagrams for the five 
frequencies listed in Table~\ref{tab2}. 
From top to bottom the data has been phased using frequencies 
$f_1$ to $f_5$)} 
\label{fasato}
\end{figure}

Finally, we performed the Fourier analysis of the whole data set. 
This procedure identifies five frequencies that satisfy the above 
requirement. The periodograms are shown in Fig.~\ref{fig3}, 
where the solid line  displays the S/N=4 level, and the 
dotted and dashed lines represent the 99\% and 90\%  
significance levels of the Scargle test, respectively.
The derived frequencies, amplitudes, phases and relative S/N 
are listed in Table~\ref{tab2}; the corresponding phase diagrams are 
shown in Fig.~\ref{fasato}. A comparison between the 
frequencies extracted from the whole data set and those
obtained during the single winter 2001 and 2002 campaigns 
is given in Table~\ref{campagne}. 
Inspection of the table reveals that frequencies $f_1$ to $f_4$ have 
been detected within the errors both in the winter 2001 and 2002 campaigns,
whereas $f_5$ is present only in the winter 2002 campaign. 

We note that in Fig.~\ref{fig3} there is another apparently 
significant frequency at 3.347 c/d with S/N$\sim$3.9.
However, we do not consider it relevant to our discussion because the period 
of $\sim$ 7.2 hr is just the duration of the longest time series obtained 
at a single site (see Sect.~\ref{obsred}), and 
the resulting frequency could be a spurious artifact. 
Whether or not this is the case goes beyond the interest of this paper,
since its inclusion in the fit has negligible influence 
on the results for the other frequencies, and in any case the frequency
is much shorter relative to the typical values of $\delta$ Scuti stars.


In order to test the Period98 results, 
we have independently carried out a frequency analysis using 
the Starlink Period package (star-www.rl.ac.uk). In particular, 
we used the Lomb Normalized Periodogram (LNP, Lomb 1976; Scargle
1982), noting that other algorithms, such 
as CLEAN  (also included in Period) give results similar 
to those obtained with LNP.

The result of this analysis is shown in the third 
column of Table~\ref{tab_starlink}, labelled LNP (1). 
There is an obvious discrepancy concerning 
$f_5$, which is not detected by LNP. 
To check if $f_5$ is reliable or not, 
we have adopted the following procedure:
1) we analysed the time series by means of LNP; 
2) once extracted the first frequency, 
we used Period98 (instead of the Period package) for the fit 
and prewithening; 3) the residuals from the prewithening have 
been analysed again using LNP; 4) we iterated the previous steps.
The result of such a procedure is shown in the fourth column of 
(labelled as  LNP (2)) of Table~\ref{tab_starlink}. This time the 
agreement with results from Period98 is almost perfect. 
 As a conclusion, our test shows how using 
different packages could lead to somewhat different results, 
suggesting the need to use more than one method to analyse the data.

As a result of the above considerations we can conclude that the
frequencies $f_1$, $f_2$, $f_3$, and $f_4$ are established beyond any 
reasonable doubt, whereas $f_5$ is slightly less reliable.

The frequency $f_5$ deserves more discussion. In fact, it is very close 
to $f_3$, the separation being only 0.06 c/d (we recall that 
the resolution for the whole data set in 0.03 c/d). 
As noted by Breger \& Bischof (2002) 
close frequency pairs have important asteroseismological 
implications, and it is prudent to investigate whether they are real or 
not. In fact, there are many observational problems listed by Breger \& 
Bischof (2002) that can lead to a false closely spaced double modes. 
In particular, amplitude variability of a single frequency can 
resemble the presence of a close pair. 
However, from the observational point of view it is 
very difficult to discern which hypothesis 
(close pair or single mode with amplitude variations) is correct. 
To this aim, 
Breger \& Bischof (2002) illustrate a method which relies on the examination 
of amplitude and phase variation with time. However, this 
analysis requires a huge amount of observations while the data on V351~Ori 
presented here is insufficient for such a test and we simply 
warn the reader that $f_5$ might not be a real pulsation frequency.  


\begin{center}
\begin{table}
\caption[]{Frequencies extracted in order of decreasing amplitude 
from the whole data set using various methods (see text).
All frequencies in the table are in c/d. 
\label{tab_starlink}}
\begin{tabular}{cccc}
\hline
\noalign{\smallskip}
Frequency & Period98 & LNP (1)  &  LNP (2)   \\
\noalign{\smallskip}
\hline
\noalign{\smallskip}
$f_1$  &    15.687  &   15.684  &  15.684  \\
$f_2$  &    14.331  &   14.329  &  14.329  \\
$f_3$  &    12.754  &   12.754  &  12.754  \\ 
$f_4$  &    15.885  &   15.891  &  15.876  \\
$f_5$  &    12.817  &    --     &  12.795  \\
\noalign{\smallskip}
\hline
\end{tabular}
\end{table}
\end{center}




\section{Discussion}

\subsection{Comparison with previous work}
\label{confr}

Table~\ref{tabx} shows the frequencies derived in the present work, 
along with those found by M01 and BKW.
With respect to M01, we are able to
confirm within the uncertainties only 3 frequencies: 
$f$=15.49~c/d, $f$=11.89~c/d (this is likely the -1 c/d 
alias of our $f_3$), and $f$=16.27~c/d.
Concerning the other three frequencies reported in M01, 
they are likely due the noise in the observations.

\begin{center}
\begin{table}
\caption[]{Comparison with previous work. 
The labels ``(P)'' and ``(RD)'' indicate if 
the frequency has been detected on the basis of photometric or 
radial velocities data. 
An uncertain correspondence of previous frequencies with the present work
is marked by a question mark. 
The uncertainties on the frequencies are also indicated.}
\label{tabx}
\begin{tabular}{ccccc}
\hline
\noalign{\smallskip}
  & This work        & M01             &  BKW (P) &   BKW (RD) \\
  & $\pm$ 0.03 (c/d) & $\pm$ 0.6 (c/d) & $\pm$ 0.08 (c/d)  &  $\pm$ 0.08 (c/d)      \\
\noalign{\smallskip}
\hline
\noalign{\smallskip}
$f_1$  &    15.687   & 15.49  &  15.675 & 15.682   \\
$f_2$  &    14.331   &   --   &  14.335 & 14.153   \\
$f_3$  &    12.754   & 11.89? &  --     & 11.877 ? \\ 
$f_4$  &    15.885   & 16.27  &  --     & --       \\
$f_5$  &    12.813   &   --   &  --     & 11.877 ? \\
\noalign{\smallskip}
\hline
\end{tabular}
\end{table}
\end{center}

On the basis of the SAAO photometry and radial velocity measurements of V351~Ori, 
BKW have identified two and three pulsational frequencies respectively 
(see the last two column of Table~\ref{tabx}). In particular they 
found  15.675 c/d and  14.335 c/d from the photometry and  15.682 c/d,
 14.153 c/d and 11.877 c/d from radial velocities data (two of which are
 clearly the same as found in the photometry). 
BKW also performed a frequency analysis on the basis of line profile
variations (see their Sect.~6 for details) that yielded a 
frequency at 1.90 c/d, considered highly uncertain and 
found indications for an additional frequency at about 20 c/d.
Comparison with the BKW results (see Table~\ref{tabx}) shows that
we confirm the presence of the two frequencies at 15.675/15.682 c/d and
14.335/14.153 c/d (our $f_1$ and $f_2$). The third feature at 11.877 c/d 
could be interpreted as the $-$1 c/d alias of our $f_3$ within 2 $\sigma$ 
or of $f_5$ within 1 $\sigma$. 
Alternatively, since our 1 c/d alias is not negligible, our 
frequency $f3$ could be the alias and the frequency at 11.877 c/d the 
correct one. An additional hypothesis is that this 
frequency does not correspond neither to $f_3$ nor to $f_5$, i.e. it 
is an additional pulsational frequency which shows up
only with spectroscopic measurements.
Concerning the remaining two periodicities reported by BKW, 
we do not find any power at 20 c/d and
we cannot say much on the frequency at 1.90 c/d, 
because of our reduction technique.

\subsection{Theoretical constraints}

Following the same approach adopted in M01, we have
tried to reproduce the five $\delta$ Scuti-like frequencies derived in
the previous section using linear non-adiabatic models of PMS $\delta$
Scuti stars.  For each selected mass, the model luminosity and effective
temperature were chosen from PMS evolutionary calculations and the 
constraint is provided by the observed pulsation period(s).
Note that the $f_1$ and $f_4$ frequencies cannot be reproduced simultaneously 
since their separation is too small and comparable to the theoretical 
uncertainty due to the model mass zoning. The same holds for $f_3$ and $f_5$, 
but, as pointed out in the previous section, the latter frequency is slightly 
less reliable. 

\begin{figure}
\psfig{figure=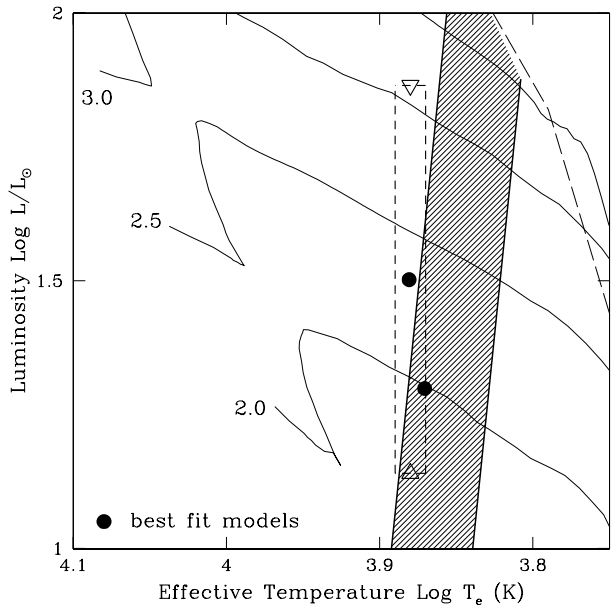,width=8.5cm}
\caption{Location in the HR diagram of the best fit models for V351~Ori 
(filled circle). The box (dashed lines and open triangles) 
indicates the range of luminosity and effective 
temperature corresponding to the empirical estimates
available in the literature. Also shown are the evolutionary
tracks by Palla \& Stahler (1993) (solid lines), and the
instability strip by Marconi \& Palla (1998) (shaded area)}
\label{fit1}
\end{figure}

On this basis, we have tried to reproduce only the three frequencies with the 
highest amplitude, $f_1$, $f_2$, and $f_3$.  
The results show that two radial pulsation models can be 
found that simultaneously pulsate with frequencies $f_1$ and $f_3$, 
corresponding to a mixed oscillation either in the first and second overtone
or in the second and third overtone. 
We caution that at present we cannot indicate which of the models is the
real one.
The stellar parameters of the two {\it best fit} 
pulsation models are given in Table~\ref{fit}
 and the position in the HR diagram is displayed in Fig.~\ref{fit1} (filled
circles).  The dashed box indicates the range in 
luminosity and effective temperature corresponding to the empirical 
estimates available in the  literature.
The minimum luminosity comes from the Hipparcos lower limit to 
the distance of 210 pc (van den Ancker et al. 1998), 
whereas the upper limit assumes that V351~Ori is located
in the Orion star forming region.  The width of the box corresponds to an
uncertainty of $\pm$0.01 dex around the mean value of
$\log T_{eff}\simeq 3.88$ (van den Ancker et al.  1998).  
The position of the best
fit models indicates that the distance of V351 Ori is intermediate
between the lower limit set by Hipparcos and that of the Orion Nebula. 

Despite the success in reproducing $f_1$ and $f_3$, our radial models fail to
also account for $f_2$. This fact, and the inability
to distinguish between $f_1$ and $f_4$, and between $f_3$ and $f_5$ indicate
that V351 Ori may be pulsating in a mixture of radial and nonradial modes, 
in agreement with the early suggestion of BKW.
The possibility that nonradial modes might be associated with the identified 
frequencies clearly requires more theoretical investigation.
Preliminary tests based on the comparison between observed and predicted
large frequency separations (by means of scaling laws) seem to suggest that the 2$M_{\odot}$
 best fit model  can be compatible with the observations if
 we consider that
$f_1$ and $f_3$ are $l=0$ modes of consecutive mode order, whereas $f_2$ is 
a $l=1$ mode. A similar test for the 2.3$M_{\odot}$ model suggests that in this 
case  it is much more difficult to reconcile model predictions
 with the observed frequencies and frequency separations.
However, these results, based on scaling considerations, need to be confirmed 
through the computation of specific nonradial models and we plan to address
this issue in detail in a forthcoming paper. 

\begin{table}
\caption[]{Stellar parameters from the best-fit models}
\label{fit}
\begin{tabular}{ccccl}
\hline
\noalign{\smallskip}
Frequency & Radial mode & $M_\ast$ & $\log L_\ast$ & $T_{eff}$ \\
     & & ($M_\odot$) & ($L_\odot$) &  (K)  \\
\noalign{\smallskip}
\hline
\noalign{\smallskip}
  $f_3$ & first overtone & 2.0 & 1.30 & 7425 K \\
  $f_1$ & second overtone &  &  &  \\
 & & & & \\
  $f_3$  & second overtone & 2.3 & 1.50 & 7600 K \\
  $f_1$  & third overtone &  &  &  \\
\noalign{\smallskip}
\hline
\end{tabular}
\end{table}

\section{Conclusions}

A total of 180 hour of observations obtained during 29 nights at 
seven different telescopes on the PMS $\delta$ Scuti star
V351~Ori have been presented. The Fourier analysis 
of this data set confirms the multiperiodic nature of
this pulsator: we have identified five frequencies of 
pulsation, four of which are highly significant, while the last is 
more uncertain even if it is  significant on the basis 
of both the Scargle test and the Breger Criterion. \par

The frequencies presented in this paper differ from those 
reported in M01, implying that the theoretical interpretation presented
there needs to be revised. 
 The comparison with the predictions of linear
nonadiabatic radial pulsation models, in the hypothesis that  
the mode identification is retained, suggests that the distance
to V351 Ori is intermediate between the lower limit measured by Hipparcos
and the distance to the Orion Nebula. However, the fact that
radial pulsation models are 
not able to reproduce simultaneously all the observed frequencies
strongly suggests that nonradial modes might also be present. 
We defer a complete analysis of the pulsation modes to a forthcoming study.

\begin{acknowledgements}
We are indebted to L. Balona, C. Koen and F. van Wyk for 
having provided us with their data in advance of publication 
and in particular to C. Koen for a critical reading of the manuscript.\par
We acknowledge R. Silvotti and L. Piau for  helpful 
discussions. We wish also to thank the Loiano staff members 
for their kind support during the observations. \par
AAF is thankful to DGAPA-UNAM project IN113599 for financial support.
FJGP acknowledges support from project ESO/FNU/ 43658/2001. DFMF
acknowledges financial support by FCT from the ``Subprograma Ci\^encia e
Tecnologia do 3$^o$ Quadro Comunit\'ario de Apoio'' and also from
project POCTI/1999/FIS/34549, approved by FCT and POCTI, with funds from
the European Union programme FEDER.\par
This work was partly supported
by FCT-Portugal, through project POCTI/43658/FNU/2001. \par
TDO would like to acknowledge partial support for this project from
NASA (grant NAGW5-9408) and NSF (AST-0097616).
\end{acknowledgements}

\end{document}